\documentclass[iop]{emulateapj}

\usepackage[bookmarks,breaklinks]{hyperref}
   \hypersetup{
     colorlinks,
     citecolor=blue,
     linkcolor=blue,
    }

\newcommand{\red}[1]{\textcolor{black}{#1}}

\shorttitle{Insight-HXMT observations of BHC MAXI J1535-571}
\shortauthors{HUANG et al.}

\received{}
\revised{}
\accepted{}

\begin{document}

\title{\emph{Insight}-HXMT observations of the New Black Hole Candidate MAXI J1535-571: timing analysis}

\author{Y. Huang$^{1,2}$, J. L. Qu$^{1}$, S. N. Zhang$^{1,2}$, Q. C. Bu$^{1}$,
Y. P. Chen$^{1}$, L. Tao$^{1}$, S. Zhang$^{1}$, F. J. Lu$^{1}$, T. P. Li$^{1,2,3}$,
L. M. Song$^{1}$
% ...
\\
(The Insight-HXMT Collaboration)}

\altaffiltext{1}{Key Laboratory of Particle Astrophysics, Institute of High Energy Physics, Chinese Academy of Sciences, Beijing 100049, China}
\altaffiltext{2}{University of Chinese Academy of Sciences, Chinese Academy of Sciences, Beijing 100049, China}
\altaffiltext{3}{Department of Physics, Tsinghua University, Beijing 100084, China}
\altaffiltext{4}{Department of Astronomy, Beijing Normal University, Beijing 100088, China}
\altaffiltext{5}{Department of Engineering Physics, Tsinghua University, Beijing 100084, China}
\altaffiltext{6}{Institute of Astronomy, National Tsing Hua University, Hsinchu 30013, Taiwan}
\altaffiltext{7}{Department of Physics, National Tsing Hua University, Hsinchu 30013, Taiwan}
\altaffiltext{8}{Key Laboratory for Research in Galaxies and Cosmology, Shanghai Astronomical Observatory, Chinese Academy of Sciences, Shanghai 200030, China}
\altaffiltext{9}{Department of Physics and Institute of Theoretical Physics, Nanjing Normal University, Nanjing 210023, China}
\altaffiltext{10}{Yunnan Observatories, Chinese Academy of Sciences, Kunming 650216, China}
\altaffiltext{11}{Key Laboratory for the Structure and Evolution of Celestial Objects, Chinese Academy of Sciences, Kunming 650216, China}
\altaffiltext{12}{Center for Astronomical Mega-Science, Chinese Academy of Sciences, Beijing 100012, China}

\email{qujl@ihep.ac.cn, zhangsn@ihep.ac.cn}

\begin{abstract}

We present the X-ray timing results of the new black hole candidate (BHC)
MAXI J1535-571 during its 2017 outburst from Hard X-ray Modulation Telescope (\emph{Insight}-HXMT) observations
taken from 2017 September 6 to 23. Following the definitions given by \citet{Belloni2010},
we find that the source exhibits state transitions from Low/Hard state (LHS) to
Hard Intermediate state (HIMS) and eventually to Soft Intermediate state (SIMS).
Quasi-periodic oscillations (QPOs) are found in the intermediate states,
which suggest different types of QPOs. With the large effective area of \emph{Insight}-HXMT
at high energies, we are able to present the energy dependence of
the QPO amplitude and centroid frequency up to 100 keV which is rarely explored
by previous satellites. We also find that the phase lag at the type-C QPOs
centroid frequency is negative (soft lags) and strongly correlated with the centroid
frequency. By assuming a geometrical origin of type-C QPOs,
the source is consistent with being a high inclination system.

\end{abstract}


\section{Introduction} \label{sec:intro}

Black hole transients (BHTs) spend most of their lives in quiescence, and are detected during outbursts
in which their spectral and timing properties change with time. During a typical outburst,
they go through the low hard state (LHS), the hard and soft intermediate states (HIMS, SIMS),
high soft state (HSS), then again through the intermediate states and back to the LHS, following
the classification given in Belloni (\citeyear{Belloni2010}, and see \citealt{McClintockRemillard2006}
for an alternative classification, and \citealt{MottaBelloniHoman2009} for a comparison). In the LHS,
the X-ray spectrum can be approximately described by a power-law with a spectral index of $\sim$ 1.6 (2-20 keV band),
and an exponential cutoff at $\sim$ 100 keV. This hard X-ray emission is thought to arise from the Comptonization
of soft disk photons in a hot corona. The corresponding power density spectrum (PDS) shows
strong ($\sim$30\% rms) band-limited noise, and sometimes low-frequency quasi-periodic oscillations (LFQPOs).
While the X-ray spectrum in the HSS is dominated by a soft thermal component, modeled with a multi-temperature disk-blackbody
with a typical temperature of $\sim$1 keV at inner disk radius, its PDS shows weak (down to few percent fractional rms) power-law noise.

Compared to the two main states, which show consistent behaviors, other states are complex and more difficult to classify
and to interpret; both disk and power-law components are clearly present in the energy
spectra, and the main feature of PDS is LFQPOs with centroid frequency ranging from a
few mHz to $\sim$30 Hz. Several types of LFQPOs have been identified and classified
into type A, B, C \citep{Remillard2002, Casella2005}. The study of LFQPOs is essential
to our understanding of the accretion flow around black holes, though their origin is still
in debate. One of the promising models for type-C QPO is that the oscillations are produced by
the Lense-Thirring precession of the inner accretion flow \citep{2009MNRAS.397L.101I}.
Evidence in support of such scenario is inferred from \red{the modulation of the reflected iron
line equivalent width \citep{2015MNRAS.446.3516I} and the centroid energy \citep{2016MNRAS.461.1967I}
during a QPO cycle by using phase-resolved spectroscopy of type-C QPO},
the inclination dependence of QPO phase lags \citep{2017MNRAS.464.2643V} and absolute
variability amplitude \citep{2015MNRAS.447.2059M}. In addition, it is also important to
consider the energy dependence of the QPO properties, such as fractional rms, centroid
frequency and time-lag \citep{2001ApJ...548..401T,2004ApJ...615..416R,2010ApJ...710..836Q,2016ApJ...833...27Y}.
It can bridge over the energy spectra and the timing variability.

The new X-ray transient, MAXI J1535-571, was independently discovered
by \emph{MAXI}/GSC \citep{2017ATel10699....1N}and \emph{Swift}/BAT \citep{Swift2017GCN, 2017ATel10700....1K}
on September 02, 2017 (MJD 57998). The radio \citep{2017ATel10711....1R},
sub-millimetres \citep{2017ATel10716....1D}, near-infrared \citep{2017ATel10716....1D}
and optical \citep{2017ATel10702....1S} counterparts were detected soon after
the discovery of the source. \emph{MAXI}/GSC and the ATCA follow-up observations indicate the source as a BHC,
judging from its X-ray spectral shape and rapid X-ray variability \citep{2017ATel10708....1N},
as well as the radio versus  X-ray luminosity ratio \citep{2017ATel10711....1R}.
Later \emph{MAXI}/GSC and \emph{Swift} observation suggested that the source was
undergoing a hard-to-soft state transition \citep{2017ATel10729....1N,2017ATel10731....1K,2017ATel10733....1P,2017submit}.
LFQPOs have been detected by \emph{Swift}/XRT and  \emph{NICER} \citep{2017ATel10734....1M,2017ATel10768....1M}.
\red{Using \emph{NICER} data, \citet{2018ApJ...860L..28M} analyzed the spectrum of MAXI J1535-571
observed on September 13. Their results gave a spin of $0.994(2)$, and a inclination angle of $67.4(8)$$^{\circ}$.}
\citet{2018ApJ...852L..34X} performed spectral fits of \emph{NuSTAR} observation using a relativistic
reflection models, and estimate a black hole spin $a > 0.84$ and a high
inclination angle: $57_{-2}^{+1}$$^{\circ}$ and $75_{-4}^{+2}$$^{\circ}$.

In this paper, we study the temporal variation of the source using \emph{Insight}-HXMT
observations. In Section 2, we describe \emph{Insight}-HXMT observations and data
reductions methods. The results are presented in Section 3. Discussions and Conclusions
follow in Section 4 and 5.

\section{Observations and Data Analysis} \label{sec:obs}

Following \emph{MAXI}/GSC and \emph{Swift}/BAT discovery of MAXI J1535-571,
we triggered \emph{Insight}-HXMT Target of Opportunity (ToO) observations.
Our follow-up observations started on September 6, 2017 and ended on
September 23, 2017, when the source was unobservable due to the Sun constraint
of the satellite. During this period, the detectors were switched off from
September 07 to 12 due to the X9.3 Solar flare
\footnote{\url{https://www.solarmonitor.org/goes_pop.php?date=20170906&type=xray}}. Our sample contains 31 pointed
observations, with each observation covering several satellite orbits.
The observation log is shown in Table \ref{tab:table1}.

\begin{deluxetable*}{cccccccc}
\tablecaption{\red{\emph{Insight}-HXMT observation of MAXI J1535-571} \label{tab:table1}}
\tablehead{
\colhead{ObsID\tablenotemark{a}} & \colhead{Start Date} & \colhead{MJD} & \colhead{obs time} & \colhead{HE rate} & \colhead{ME rate} & \colhead{LE rate} & \colhead{State\tablenotemark{b}} \\
\colhead{} & \colhead{} & \colhead{} & \colhead{(ks)} & \colhead{(cts s$^{-1}$)} & \colhead{(cts s$^{-1}$)} & \colhead{(cts s$^{-1}$)} & \colhead{} \\
\colhead{} & \colhead{} & \colhead{} & \colhead{} & \colhead{($26-100$ keV)} & \colhead{($6-38$ keV)} & \colhead{($1-12$ keV)} & \colhead{}
}
\startdata
105 & 2017-09-06 & 58002.317 & 13 & -\tablenotemark{c}  & $355\pm4$  & $299\pm1$ & LHS\\
106 &   & 58002.469 & 11 & -\tablenotemark{c}  & $365\pm4$  & $374\pm2$ & LHS\\
107 &   & 58002.601 & 11 & -\tablenotemark{c}  & $373\pm5$  & $386\pm2$ & LHS\\
108 &   & 58002.734 & 11 & $850\pm27$  & $390\pm5$  & $405\pm2$ & LHS\\
119 & 2017-09-12 & 58008.443 & 12 & $632\pm27$  & $627\pm5$  & $1447\pm2$ & HIMS\\
120 &   & 58008.583 & 38 & $623\pm28$  & $623\pm5$  & $1501\pm2$ & HIMS\\
121 & 2017-09-13 & 58009.029 & 10 & $636\pm24$  & $648\pm4$  & -\tablenotemark{d} & HIMS\\
122 &   & 58009.156 & 17 & $685\pm28$  & $672\pm5$  & -\tablenotemark{d} & HIMS\\
201 & 2017-09-14 & 58010.205 & 11 & $795\pm29$  & $742\pm5$  & -\tablenotemark{d} & HIMS\\
301 & 2017-09-15 & 58011.200 & 11 & $787\pm28$  & $770\pm5$  & $1638\pm3$ & HIMS\\
401 & 2017-09-16 & 58012.260 & 11 & $728\pm24$  & $765\pm4$  & $1873\pm2$ & HIMS\\
501 & 2017-09-17 & 58013.255 & 11 & $697\pm24$  & $788\pm4$  & $2123\pm2$ & HIMS\\
601 & 2017-09-18 & 58014.117 & 11 & $714\pm29$  & $820\pm5$  & $2208\pm3$ & HIMS\\
701 & 2017-09-19 & 58015.974 & 12 & $312\pm27$  & $522\pm6$  & $3212\pm4$ & SIMS\\
902 & 2017-09-21 & 58017.250 & 12 & $455\pm26$  & $655\pm5$  & $3303\pm2$ & SIMS\\
903 &   & 58017.389 & 12 & $486\pm35$  & $755\pm7$  & $3208\pm3$ & SIMS\\
904 &   & 58017.529 & 32 & $365\pm26$  & $613\pm5$  &  & SIMS\\
905 &   & 58017.902 & 11 & $211\pm26$  & $352\pm6$  & $3385\pm5$ & SIMS\\
906 & 2017-09-22 & 58018.032 & 12 & $237\pm30$  & $377\pm5$  & $3174\pm3$ & SIMS\\
907 &   & 58018.173 & 12 & $222\pm25$  & $369\pm5$  & $3171\pm3$ & SIMS\\
908 &   & 58018.314 & 12 & $260\pm30$  & $447\pm7$  & $3216\pm4$ & SIMS\\
909 &   & 58018.453 & 32 & $201\pm26$  & $331\pm5$  &  & SIMS\\
910 &   & 58018.832 & 10 & $192\pm25$  & $342\pm7$  & $3408\pm5$ & SIMS\\
911 &   & 58018.958 & 11 & $195\pm29$  & $339\pm6$  & $3192\pm3$ & SIMS\\
912 & 2017-09-23 & 58019.093 & 12 & $216\pm26$  & $356\pm5$  & $3175\pm3$ & SIMS\\
913 &   & 58019.238 & 12 & $237\pm26$  & $383\pm6$  & $3159\pm3$ & SIMS\\
914 &   & 58019.377 & 12 & $201\pm29$  & $352\pm6$  & $3147\pm3$ & SIMS\\
915 &   & 58019.517 & 12 & $265\pm26$  & $415\pm5$  & $3194\pm3$ & SIMS\\
916 &   & 58019.657 & 11 & $273\pm27$  & $477\pm6$  & $3234\pm3$ & SIMS\\
917 &   & 58019.789 & 11 & $264\pm27$  & $462\pm5$  & $3311\pm4$ & SIMS\\
918 &   & 58019.921 & 9 & $261\pm28$  & $411\pm7$  & $3123\pm3$ & SIMS
\enddata
\tablenotetext{a}{105: P011453500NNN, NNN=105}
\tablenotetext{b}{Follows definitions in \citet{Belloni2010}.}
\tablenotetext{c}{HE detector was operated in the GRB mode, where the high voltage of PMT was reduced.}
\tablenotetext{d}{LE detector was saturated through this observation.}
\end{deluxetable*}

The Hard X-ray Modulation Telescope (HXMT, also dubbed as \emph{Insight}-HXMT)\citep{2014SPIE.9144E..21Z},
the first Chinese X-ray astronomical satellite, consists of three slat-collimated instruments:
the High Energy X-ray Telescope (HE), the Medium Energy X-ray Telescope (ME), and the Low
Energy X-ray Telescope (LE). HE contains 18 cylindrical NaI(Tl)/CsI(Na) phoswich detectors
which are sensitive in the 20-250 keV with a total detection area of about 5000 cm$^2$;
ME is composed of 1728 Si-PIN detectors which are sensitive in the 5-30 keV with a total
detection area of 952 cm$^2$; and LE uses Swept Charge Device (SCD) which is sensitive
in 1-15 keV range with a total detection area of 384 cm$^2$. There are three types of Field of View (FoV)
: $1^{\circ} \times 6^{\circ}$ (FWHM,full-width half-maximum) (also called the small FoV),
$6^{\circ} \times 6^{\circ}$ (the large FoV), and the
blind FoV used to estimate the particle induced instrumental background. Since its launch,
\emph{Insight}-HXMT went through a series of performance verification tests by observing
blank sky, standard sources and sources of interest. These tests \red{showed} that the
entire satellite works smoothly and healthily, and have allowed for the calibration and
estimation of the instruments background.

We use the \emph{Insight}-HXMT Data Analysis software (HXMTDAS) v2.0\footnote{\url{http://www.hxmt.org/index.php/dataan}}
to analyze all the data, filtering the data with the following criteria:
(1) pointing offset angle $< 0.05^{\circ}$; (2) elevation angle $> 6^{\circ}$;
(3) the value of the geomagnetic cutoff rigidity $> 6$. We only select events that
belong to the small FoV. Since LE detector can be saturated due to the bright
earth and local particles, we need to create the good time intervals (GTIs) manually.
For some observations there is no GTI for LE detector. Since the detailed background
model is still in progress, we use the blind FoV detectors to estimate the \emph{Insight}-HXMT background,
with a systematic error of $10\%$. We derive the background as $B = N*C_b$, where $B$ is the background counts
rate of the small FoV in a given energy band, $N$ is the ratio of number of the small FoV detectors
to that of the blind FoV detectors, and $C_b$ stands for the blind FoV detectors count rate in the same energy
band as $B$. Using blank sky observations, we tested the reliability of this method.

To study the variability, we produce the PDS from 64s data intervals with time resolution of 1/128s
for each observation; in a few cases, an inspection of the PDS show significant variations
in the QPO frequency between different orbits, which were therefore split.
The PDS is applied Miyamoto normalisation \citep{1991ApJ...383..784M} after
subtracting the Possion noise. PDS is fitted with a combination of
Lorentzians \citep{2000MNRAS.318..361N,2002ApJ...572..392B} using the XSPEC v12.9.1
between 0.01 Hz and 32 Hz. The best-fit reduced $\chi^2$ values are less than 1.5 (for a degree-of-freedom of $\sim$138 ),
with a typical value of 1.2. We estimate the total fractional variability (rms of PDS) in the range of 0.1 to 32 Hz.

We also produce 16s cross spectrum between the 1-3 keV and 3-7 keV light
curves of \emph{Insight}-HXMT/LE (defined as $C(j)=X_1^*(j)X_2(j)$, where $X_1$ and $X_2$ are
the complex Fourier coefficients for the two energy bands at a frequency $v_j$ and $X_1^*(j)$ is
the complex conjugate of $X_1(j)$), and calculate average cross spectrum vectors for each observation.
The phase lag at frequency $v_j$ is $\o_j=arg[C(j)]$. The error in $\o_j$ is computed from the
observed variance of $C$ in the real and imaginary directions. For phase lag spectra,
positive lag values mean that the hard photons are lagging the soft ones.
To quantify the phase-lag behaviour of the QPOs, we compute their phase lags in a range
centered at the QPO centroid frequency and spread over the width of the QPO \citep{2000ApJ...541..883R}.

No application of dead time correction is given in the PDS and the cross spectrum,
since dead time should not be an issue in our analysis. In \emph{Insight}-HXMT, dead time ($\tau_d$) is
around 20 us for HE and LE; 250 us for ME, thus the frequency range commonly analyzed in BHC is well below $1/\tau_d$.

\section{Results} \label{sec:results}

\subsection{Fundamental diagrams} \label{subsec:evolution}

\begin{figure*}
\plottwo{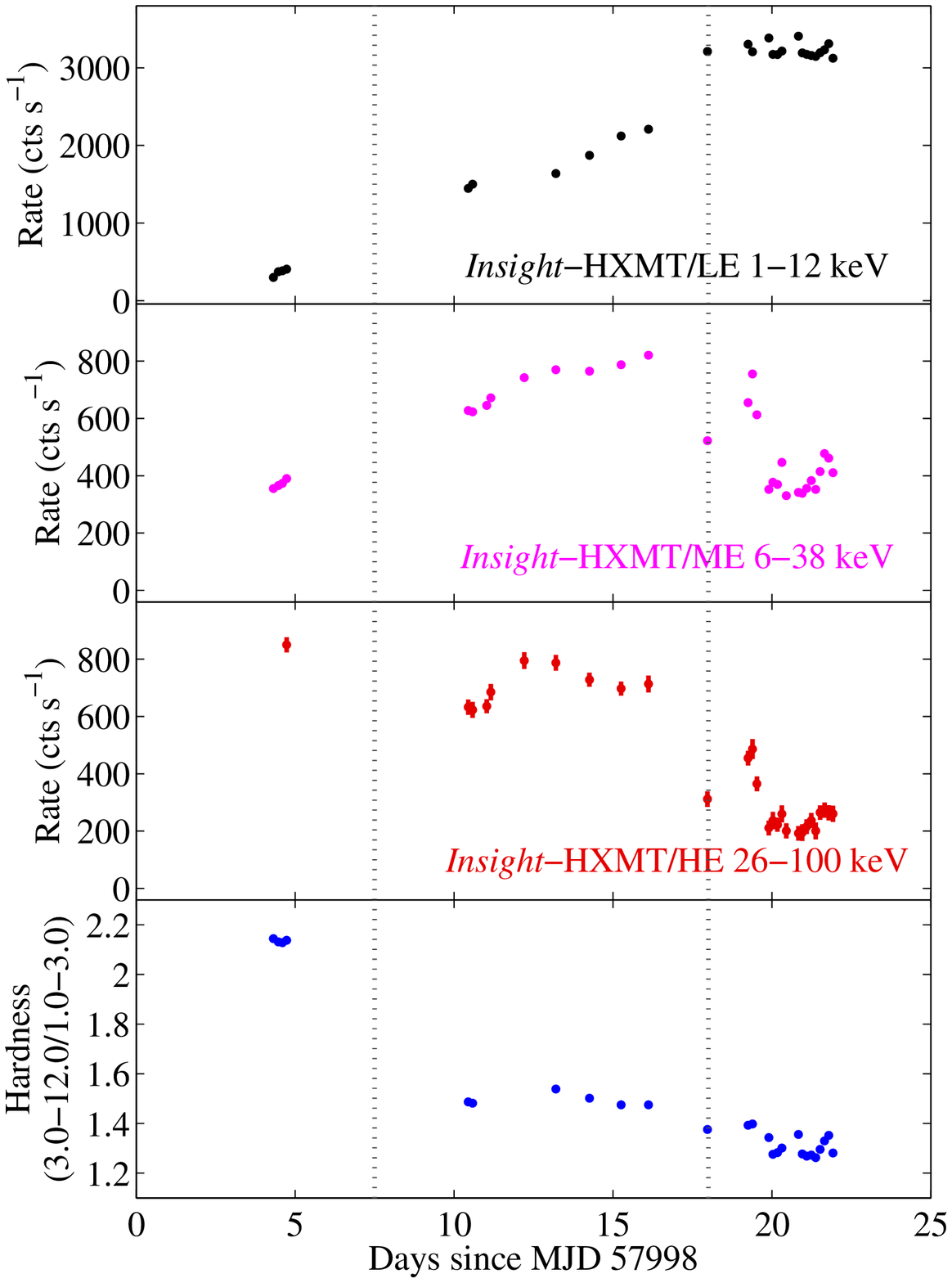}{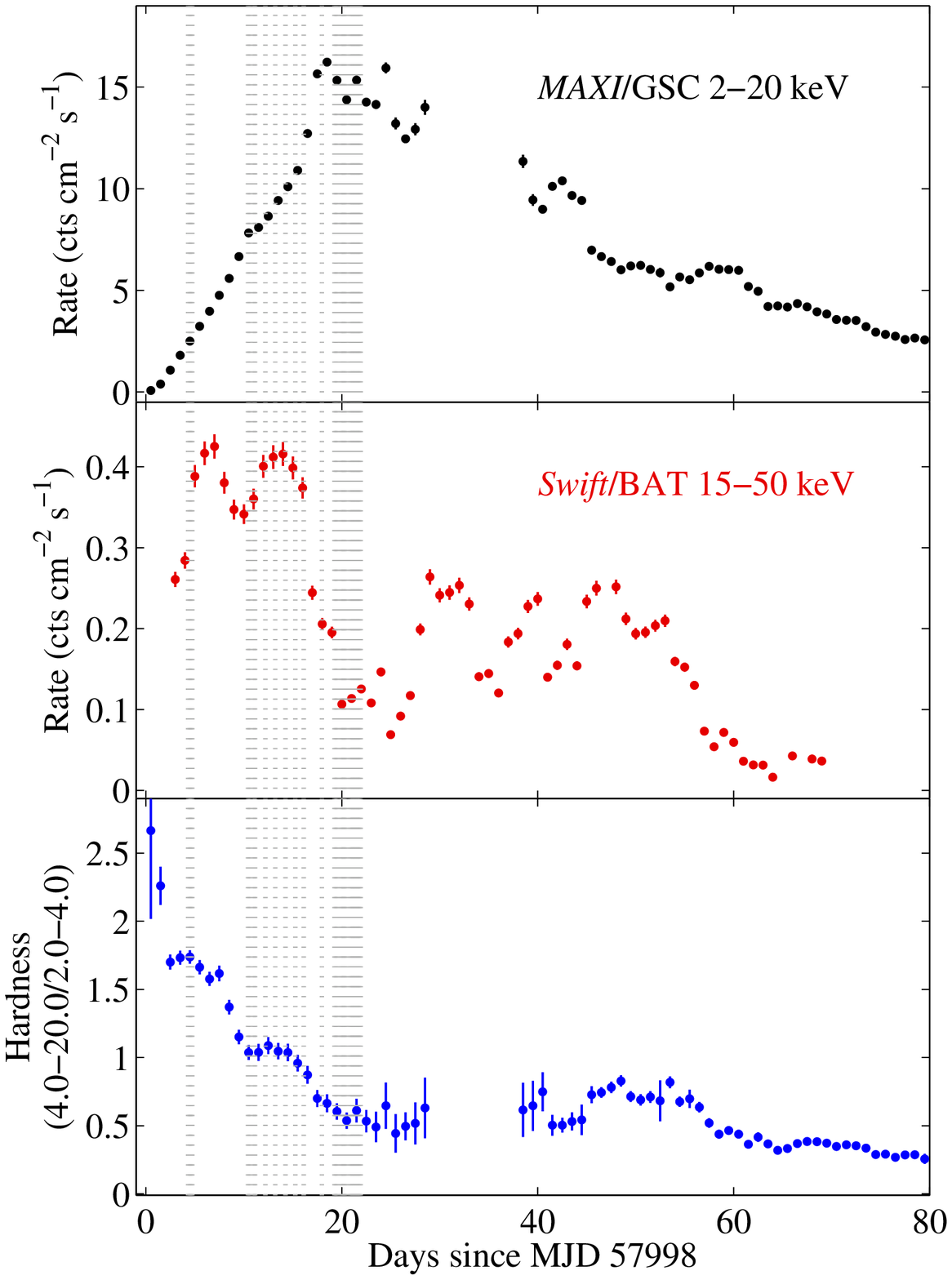}
\caption{Left Panel: the \emph{Insight}-HXMT/LE (1-12 keV), the \emph{Insight}-HXMT/ME (6-38 keV) and the
\emph{Insight}-HXMT/HE (20-90 keV) light curve, and the hardness ratio between the
LE hard energy band (3-12 keV) and soft energy band (1-3 keV) of MAXI J1535-571. Each point
represents one \emph{Insight}-HXMT observation. The vertical dashed lines indicate the transition
of states. Right Panel: \emph{MAXI}/GSC light curve, \emph{Swift}/BAT light curve and \emph{MAXI}
hardness ratio (4-20/2-4 keV) of MAXI J1535-571. The gray shaded areas mark the \emph{Insight}-HXMT
observations. \label{fig:f1}}
\end{figure*}

\begin{figure*}
\plottwo{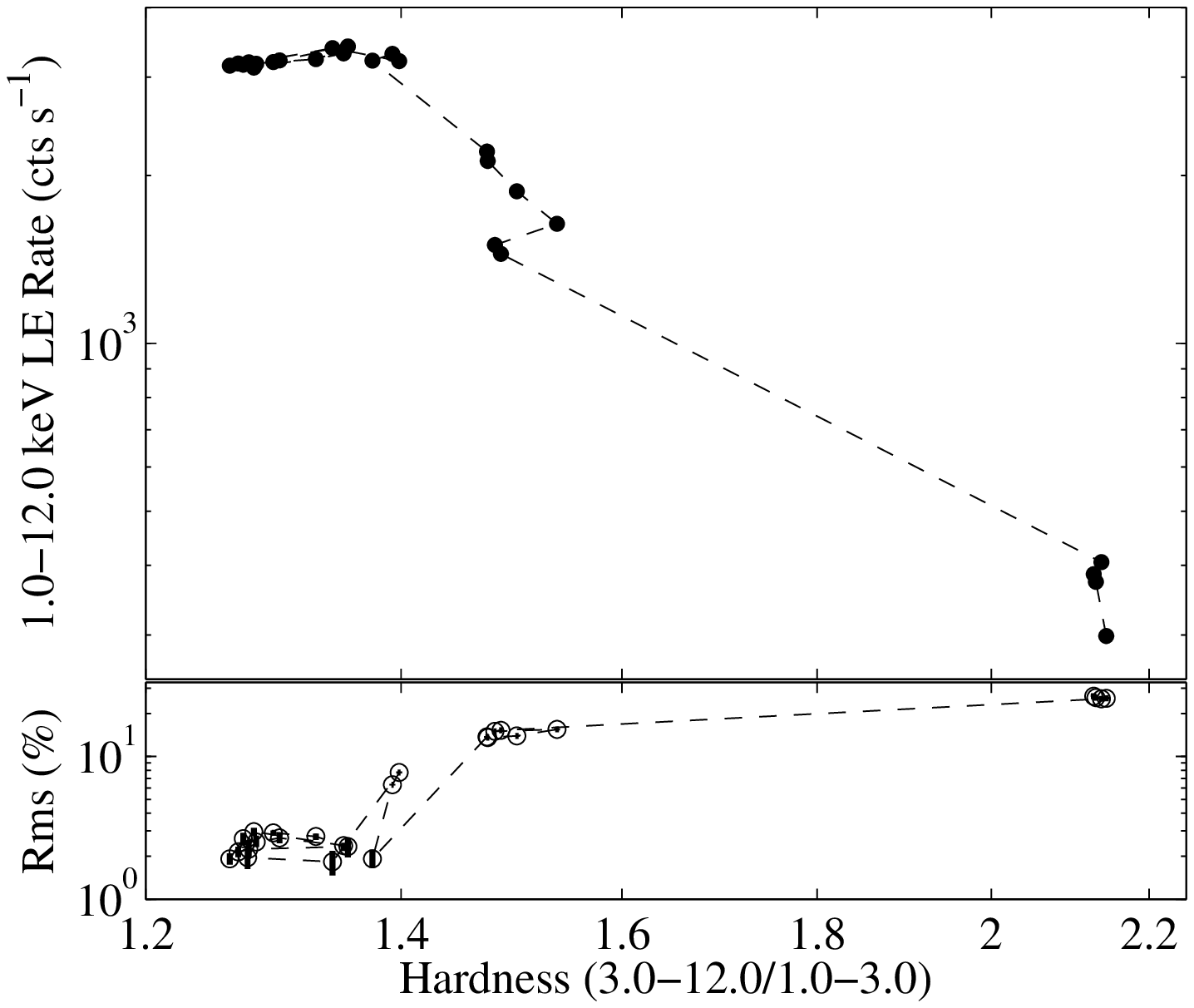}{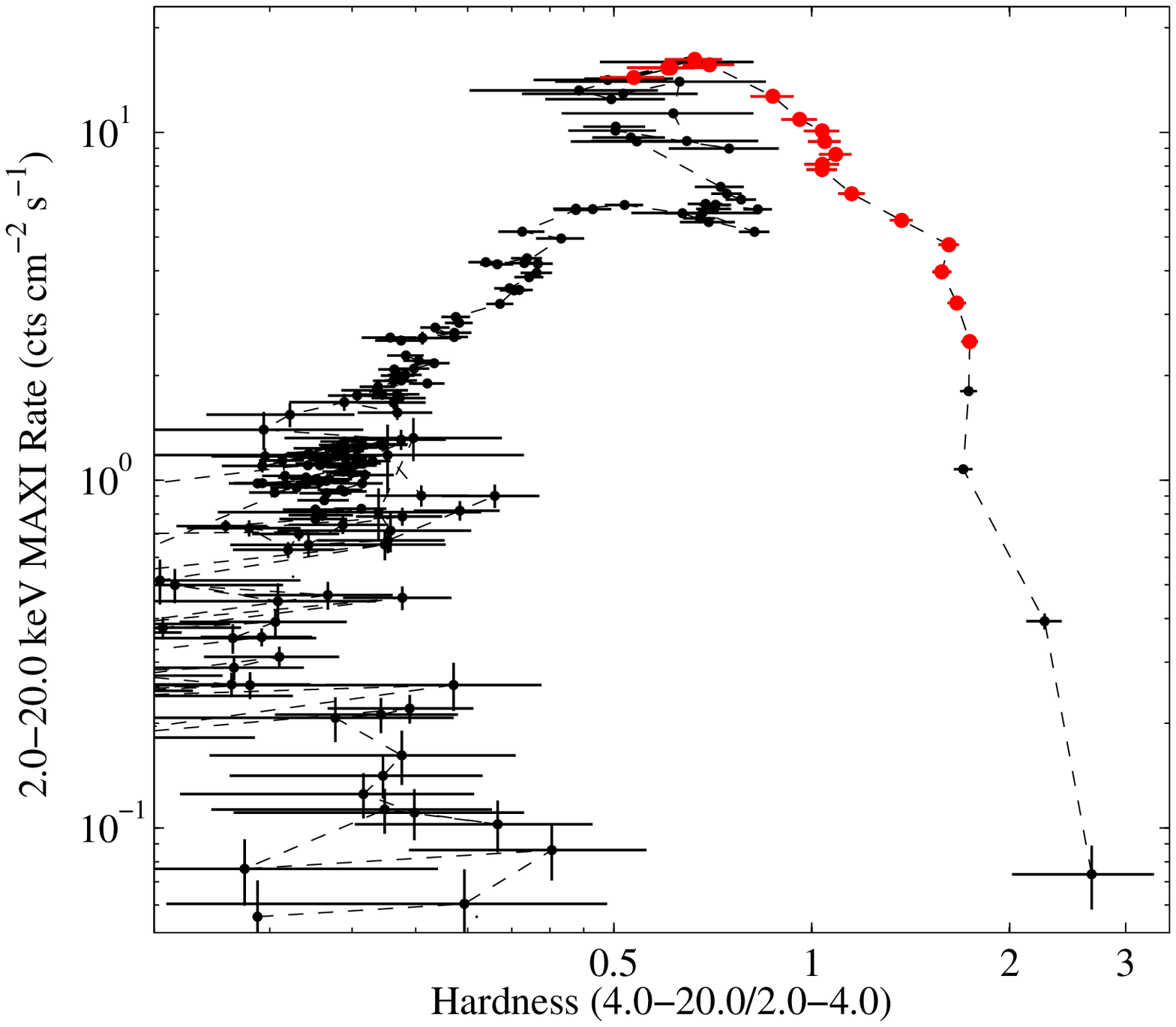}
\caption{\emph{Insight}-HXMT hardness-intensity diagram (HID), hardness-rms diagram (HRD)
and \emph{MAXI}/GSC hardness-intensity diagram of MAXI J1535-571. Left panel:
\emph{Insight}-HXMT HID (upper) and HRD (lower). Intensity is the LE count rate
in the 1.0-12.0 keV. Hardness is defined as the ratio of count rate between 3.0-12.0
keV and 1.0-3.0 keV. Fractional averaged rms corresponds to the frequency range
0.1-32 Hz to the full energy range. Each point corresponds to one observation.
Right panel: \emph{MAXI}/GSC HID. In the right panel, intensity here is the count rate in the 2.0-20.0 keV,
while hardness is defined as 4.0-20.0 keV to 2.0-4.0 keV counts ratio. The red points
indicate the time interval during which \emph{Insight}-HXMT observations were taken. \label{fig:f2}}
\end{figure*}

We plot the diagrams commonly used for the study of BHT in Figs.\ref{fig:f1} and \ref{fig:f2}.
To make a comparison, we also show the
\emph{MAXI}/GSC\footnote{\url{http://maxi.riken.jp/star\_data/J1535-572/J1535-572.html}}
and \emph{Swift}/BAT\footnote{\url{https://swift.gsfc.nasa.gov/results/transients/weak/MAXIJ1535-571/}}
results taken from the web sites for each instrument.

The background-subtracted and dead time corrected \emph{Insight}-HXMT light curves and
hardness of MAXI J1535-571 are shown in Fig.\ref{fig:f1} (left panel). The LE count rate (1-12 keV)
slowly rose from the beginning, reached its peak of 3212 cts s$^{-1}$ on MJD 58015,
and then stayed stable at that level. The ME count rate (6-38 keV) increased from
355 cts s$^{-1}$ on MJD 58002 to 820 cts s$^{-1}$ on MJD 58014, and decreased abruptly
to 522 cts s$^{-1}$ on MJD 58015, followed by several rises and falls. The HE count rate (26-100 keV)
\red{showed} a decrease in the early phase, then is similar to the ME. The hardness defined
as the count rate in the 3-12 keV energy band divided by the count rate in the 1-3 keV
energy band. We found that the hardness remained the same ($\sim$2.1) in the first several exposures around MJD $\sim$58002, but suddenly
decreased to $\sim$1.5 on MJD $\sim$58008, and then slowly decreased to a low level.
The trend of light curves and hardness observed by \emph{Insight}-HXMT, \emph{MAXI}/GSC and \emph{Swift}/BAT
are consistent with each other.

The hardness-intensity diagram (HID) and the hardness-rms diagram (HRD) are shown
in the left panel of Fig.\ref{fig:f2}. Because only the rising part of the outburst
was observed by the \emph{Insight}-HXMT, the source exhibited part of the standard
q-shaped pattern. A relatively complete pattern is described by \emph{MAXI} data in the right panel,
with \emph{Insight}-HXMT observations marked with red points. The outburst starts at the lower right
of the figure, corresponding to the LHS, where the fractional rms remains at $\sim$26\%.
When the intensity increases, the source on the HID starts moving to the upper left,
and the fractional rms drops to $\sim$15\% on MJD 58008. In the corresponding PDS,
strong type-C QPOs are detected (see \S\ref{subsec:pds}), indicating that
the system is in the HIMS. It is not possible to decide the precise transition position
from the \emph{Insight}-HXMT observations, as the instruments were switched off during
that period. After several days in the HIMS, the fractional rms suddenly decreases to 1.9\%
on MJD 58015, and type-B QPOs (see \S\ref{subsec:pds}) are seen in the PDS,
indicating the system is in the SIMS. Then, the source moved irregularly in the HID but
remained in the upper left. The fractional rms increases to 7.7\%, then decreases to $\sim$2\%.

\subsection{Power Density Spectra} \label{subsec:pds}

\begin{figure}
\plotone{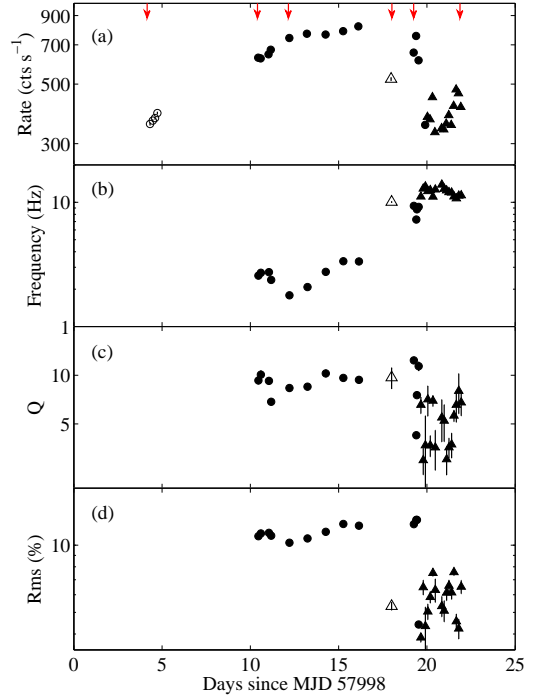}
\caption{a) \emph{Insight}-HXMT/ME 6-38 keV light curves of MAXI J1535-571 during
the outburst. Panels b), c) and d) show the evolution of the frequency, Q value,
rms of the QPO with time. The red arrows in Panel (a) indicate 6 representative
observations for which the power spectra are plotted in Fig.\ref{fig:f3}.
The open circles, filled circles, open triangles, and filled triangles
denote observations which show no QPOs, type-C QPOs, type-B QPOs and type-A QPOs, respectively\label{fig:f4}}
\end{figure}

\begin{figure*}
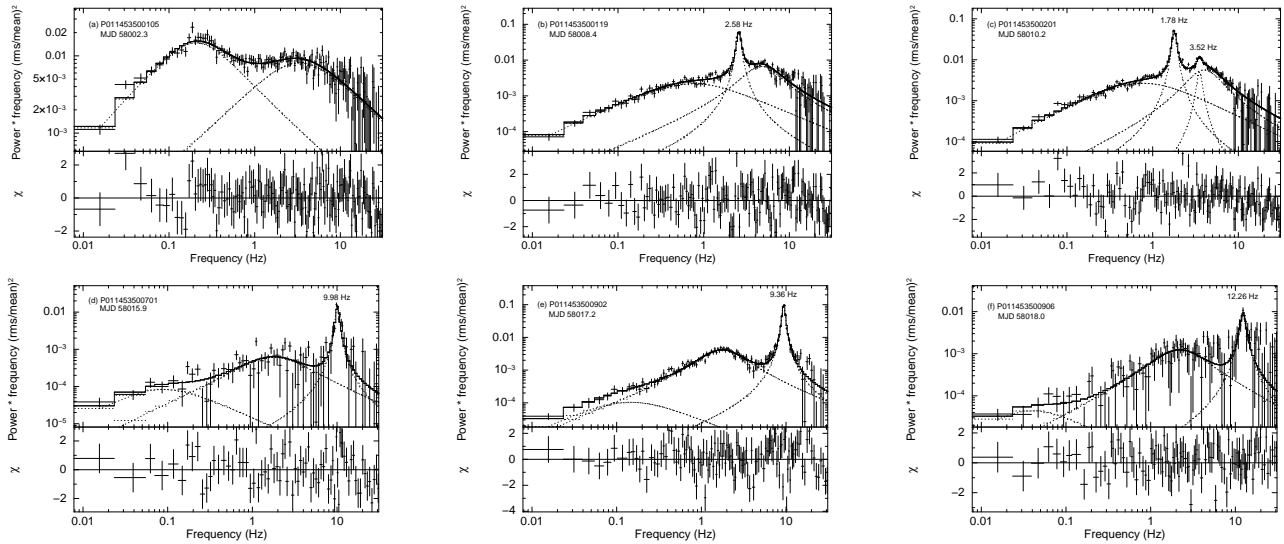

\centering
\begin{minipage}{0.32\textwidth}
\includegraphics[height=0.9\linewidth,angle=-90]{pds_P011453500105.eps}
\end{minipage}
\begin{minipage}{0.32\textwidth}
\includegraphics[height=0.9\linewidth,angle=-90]{pds_P011453500119.eps}
\end{minipage}
\begin{minipage}{0.32\textwidth}
\includegraphics[height=0.9\linewidth,angle=-90]{pds_P011453500201.eps}
\end{minipage}
\begin{minipage}{0.32\textwidth}
\includegraphics[height=0.9\linewidth,angle=-90]{pds_P011453500701.eps}
\end{minipage}
\begin{minipage}{0.32\textwidth}
\includegraphics[height=0.9\linewidth,angle=-90]{pds_P011453500902.eps}
\end{minipage}
\begin{minipage}{0.32\textwidth}
\includegraphics[height=0.9\linewidth,angle=-90]{pds_P011453500906.eps}
\end{minipage}
\caption{The power density spectra (PDS) for the 6 representative observations
selected from Fig.\ref{fig:f4} using the \emph{Insight}-HXMT/ME data (6-38 keV).
The solid line shows the best fit with multi-Lorentzians function (dotted lines).
QPO fundamental and harmonics centroid frequencies are indicated. \label{fig:f3}}
\end{figure*}

\begin{figure}
\plotone{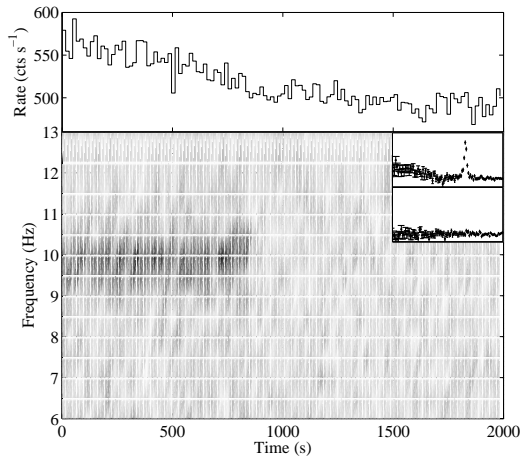}
\caption{Top panel: a 2000s segment of light curves of observation P011453500701 (16s bin).
Bottom panel: corresponding dynamical PDS, where darker points correspond to higher power. Inset:
average power spectrum from the first $\sim$800s (top) and the rest (bottom).
Count rate is 6-20 keV for ME detector. \label{fig:f12}}
\end{figure}

\begin{figure*}
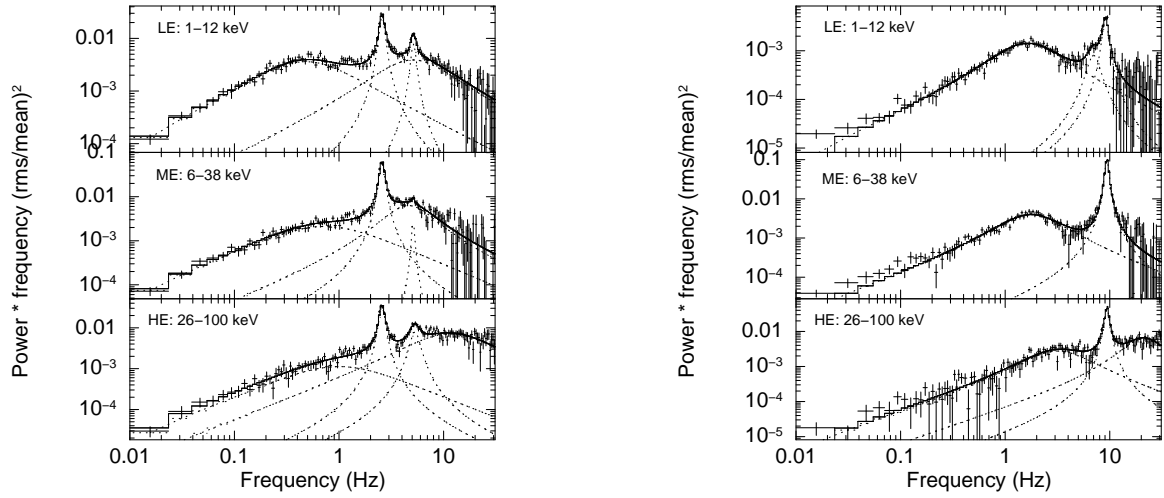

\centering
\begin{minipage}{0.48\textwidth}
\includegraphics[height=0.8\linewidth,angle=-90]{P011453500119_fig5.eps}
\end{minipage}
\begin{minipage}{0.48\textwidth}
\includegraphics[height=0.8\linewidth,angle=-90]{P011453500902_fig5.eps}
\end{minipage}
\caption{The PDS of the same observation from three detectors. The upper, middle and lower panels are for
LE: 1-12 keV, ME: 6-38 keV, HE: 26-100 keV, respectively. Left: MJD 58008.3 (ID: P011453500119).
All the PDS show the 2.58 Hz QPO, while the harmonics are more significant
in LE and HE. Right: MJD 58017.2 (ID: P011453500902).
The PDS shows that the 9.36 Hz QPO is exhibited in all the three detectors. }
\label{fig:f5}
\end{figure*}

Fig.\ref{fig:f4}-\ref{fig:f5} show results of the PDS. In Table 2, we present a summary
of the results on LFQPO parameters, i.e., the centroid frequency ($\nu$),
the coherence parameter $Q(=\nu/\Delta\nu)$ and the rms of the QPOs, $\Delta\nu$ is the FWHM of the QPO.

\begin{deluxetable}{lcccc}
\tablecaption{\red{Low-Frequency QPO Parameters for MAXI J1535-571} \label{tab:table2}}
\tablehead{
\colhead{ObsID\tablenotemark{a}} & \colhead{Type} & \colhead{QPO $\nu$\tablenotemark{b}} & \colhead{Q\tablenotemark{b}} & \colhead{rms\tablenotemark{b}} \\
\colhead{} & \colhead{} & \colhead{(Hz)} & \colhead{} & \colhead{(\%)}
}
\startdata
119 & C & $2.57\pm0.01$ & $9.3\pm0.4$ & $11.0\pm0.2$\\
120 & C & $2.71\pm0.01$ & $10.1\pm0.5$ & $11.3\pm0.2$\\
121 & C & $2.74\pm0.01$ & $9.2\pm0.4$ & $11.4\pm0.2$\\
122 & C & $2.37\pm0.01$ & $6.9\pm0.3$ & $11.0\pm0.2$\\
201 & C & $1.78\pm0.01$ & $8.3\pm0.4$ & $10.3\pm0.2$\\
301 & C & $2.08\pm0.01$ & $8.5\pm0.4$ & $10.7\pm0.2$\\
401 & C & $2.76\pm0.01$ & $10.3\pm0.5$ & $11.5\pm0.2$\\
501 & C & $3.35\pm0.01$ & $9.6\pm0.3$ & $12.4\pm0.2$\\
601 & C & $3.34\pm0.01$ & $9.4\pm0.4$ & $12.2\pm0.2$\\
701 & B & $10.06\pm0.05$ & $9.7\pm1.5$ & $5.3\pm0.2$\\
902 & C & $9.37\pm0.01$ & $12.4\pm0.3$ & $12.4\pm0.1$\\
903\_a & C & $7.28\pm0.03$ & $4.3\pm0.2$ & $13.0\pm0.2$\\
903\_b & C & $8.79\pm0.01$ & $7.5\pm0.2$ & $13.0\pm0.2$\\
904\_a & C & $9.20\pm0.02$ & $11.4\pm0.7$ & $4.4\pm0.1$\\
904\_b & A & $11.13\pm0.10$ & $6.6\pm0.8$ & $3.9\pm0.2$\\
904\_c & A & $12.9\pm0.4$ & $3.0\pm0.6$ & $6.5\pm0.5$\\
905 & A & $13.4\pm0.8$ & $4\pm3$ & $4.3\pm1.0$\\
906 & A & $12.28\pm0.11$ & $7.1\pm1.7$ & $5.0\pm0.4$\\
907 & A & $12.5\pm0.2$ & $3.7\pm0.6$ & $5.9\pm0.4$\\
908 & A & $11.10\pm0.05$ & $6.9\pm0.5$ & $7.5\pm0.2$\\
909 & A & $12.7\pm0.3$ & $3.6\pm1.2$ & $6.3\pm0.9$\\
910 & A & $13.9\pm0.3$ & $5.5\pm1.8$ & $5.3\pm0.6$\\
911 & A & $12.9\pm0.3$ & $5.2\pm1.6$ & $5.1\pm0.6$\\
912 & A & $12.5\pm0.3$ & $3.0\pm0.7$ & $6.1\pm0.5$\\
913 & A & $12.1\pm0.2$ & $3.6\pm0.5$ & $6.6\pm0.4$\\
914 & A & $12.0\pm0.2$ & $3.7\pm0.7$ & $6.1\pm0.4$\\
915 & A & $11.15\pm0.07$ & $5.6\pm0.5$ & $7.6\pm0.3$\\
916 & A & $10.76\pm0.12$ & $6.6\pm1.6$ & $4.6\pm0.3$\\
917 & A & $11.32\pm0.18$ & $8\pm3$ & $4.2\pm0.4$\\
918 & A & $11.38\pm0.10$ & $6.8\pm1.3$ & $6.5\pm0.5$
\enddata
\tablenotetext{a}{105: P011453500NNN, NNN=105}
\tablenotetext{b}{QPO centroid frequency, Q and amplitude were computed from ME detector in energy band 6-38 keV.}
\end{deluxetable}

Fig.\ref{fig:f4} shows the QPO evolution with time. In Fig.\ref{fig:f3}, we show 6 representative PDS of
\emph{Insight}-HXMT/ME whose corresponding positions are indicated by red arrows in Fig.\ref{fig:f4}.
At the beginning of the outburst (the first four exposures), the PDS shown in Fig.\ref{fig:f3}(a) is
very similar to that observed in other black holes during their typical LHSs \citep{Belloni2010},
and can be fitted with two broad Lorentzian components. Later (from MJD 58008 to MJD 58014,
Fig.\ref{fig:f3}(b)(c)), the PDS show a strong type-C QPOs, sometimes with its second harmonic,
and the centroid frequency of QPO decreased from 2.5 to 1.7 Hz and then increased to 3.3 Hz.
During the ME count rate decline on MJD 58015(see Fig.\ref{fig:f1}), we detect a $9.98$ Hz QPO
with a rather low rms amplitude (5.3\%) and a weak red-noise component at very low frequency, indicating that
 it may be type-B (Fig.\ref{fig:f3}(d)). The dynamical PDS for the first 2000 s of this observation
 showed rapid transition (see Fig.\ref{fig:f12}). During the first $\sim$800 s, the PDS showed appearances
 of type-B (with a QPO frequency $\sim$10 Hz). During the decrease phase in the light curve,
 no significant QPO with ME is detected. The HE data showed the similar behaviour, while the
 LE was saturated during this time. From MJD 58017 to the end of our sample (MJD 58023),
 the behaviour of PDS is rather complex. On MJD 58017, while the ME rate increases, we
 detected a $\sim$10 Hz QPO with a high rms amplitude ($\sim$13\%) compared to the
 previous one (Fig.\ref{fig:f3}(e)). Even though the QPO centroid frequency is
 different from previous type-C QPOs, the rms suggest that this QPO is type-C.
 After this, the QPO ($\sim$12 Hz) becomes weaker and broader with a low amplitude
 red-noise component (Fig.\ref{fig:f3}(f)), suggesting a transition to type-A QPO.

The PDS of the HE and LE detectors are
 approximately the same in the evolution. In Fig.\ref{fig:f5} we present the PDS of the three
 detectors for two observations, in which the shape of the PDS significantly evolves with energy.

\begin{figure*}
\plotone{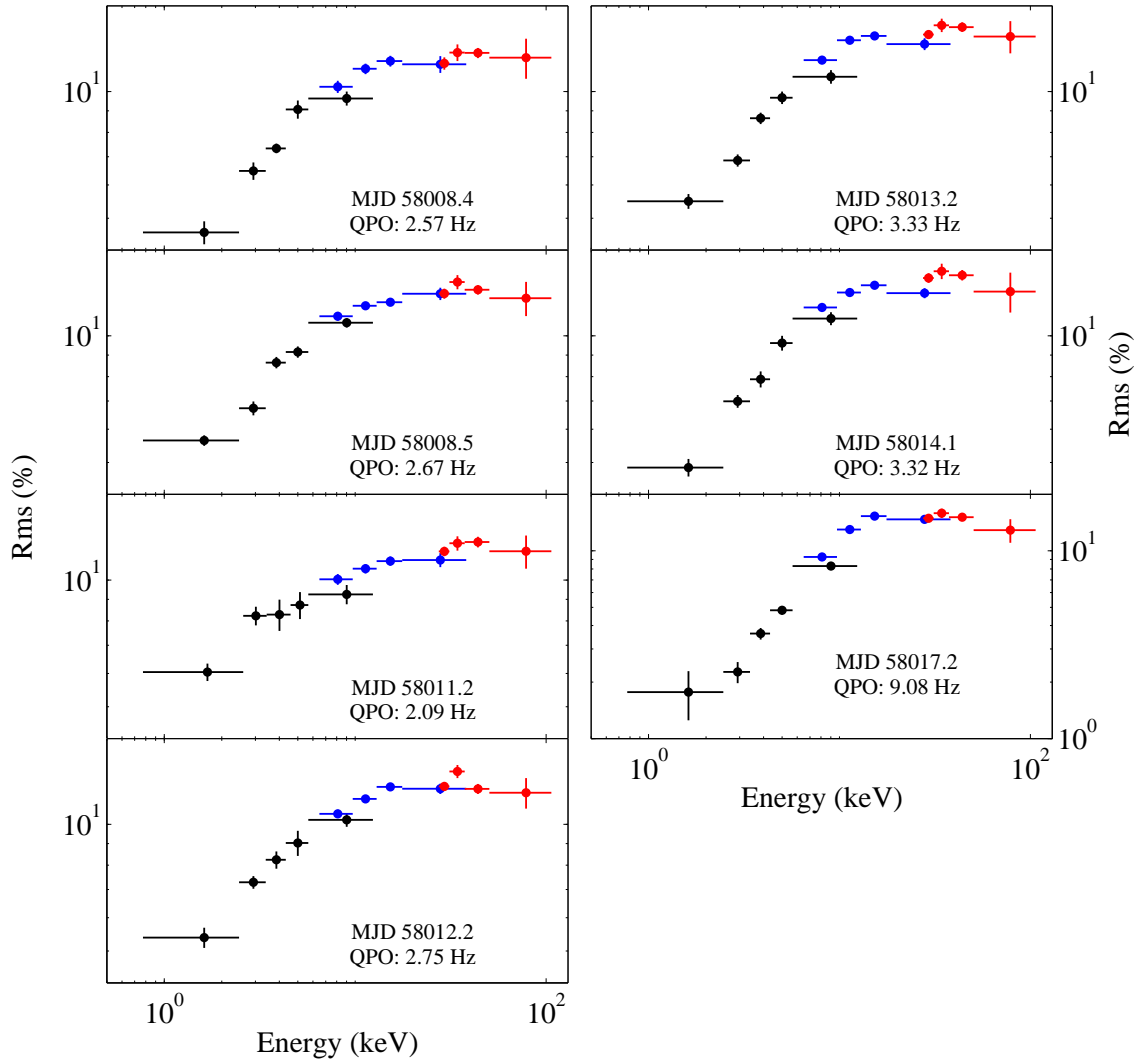}
\caption{\red{Fractional rms amplitude of the type-C QPOs as a function of photon energy.
The black, blue and red points represent LE, ME and HE data respectively.}\label{fig:f8}}
\end{figure*}

\begin{figure*}
\plotone{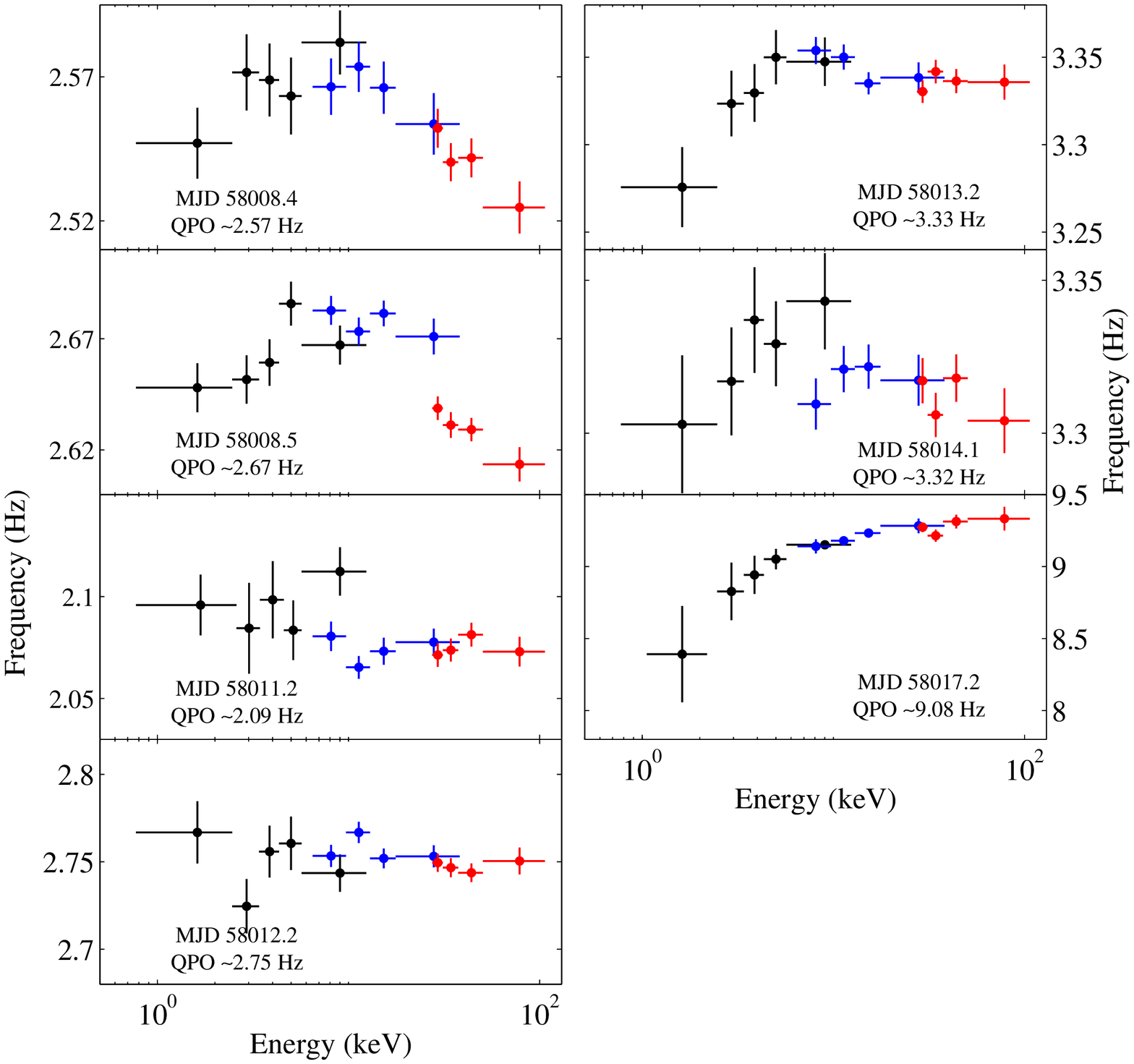}
\caption{\red{Centroid frequency of the type-C QPOs as a function of energy.
The black, blue and red points represent LE, ME and HE data respectively.}\label{fig:f6}}
\end{figure*}

In order to quantitatively study the energy dependent behavior of the QPO properties,
we extract power spectra in several energy bands. To improve the statistics,
we only derive the energy dependence of the type-C QPOs with high amplitudes.
The fractional rms and the centroid frequency of the type-C QPOs as functions
of photon energy are shown in Fig.\ref{fig:f8} and \ref{fig:f6}, with the QPO frequencies
and obsID marked in each panel. We consider the background contribution to the fractional
rms calculation. The formula is rms = $\sqrt{P}*(S+B)/S$ \citep{2015ApJ...799....2B},
where S and B stand for source and background count rates respectively, and P is the power
normalized according to Miyamoto \citep{1991ApJ...383..784M}.
In the region where LE and ME or ME and HE overlap, there is a good agreement between the
two detectors. In all cases, the rms increases with photon energy till $\sim$20 keV,
from $\sim$5\% in the lowest energy band up to $\sim$13\% during HIMS and from $\sim$1\% to $\sim$15\%
during SIMS, and stays more or less constant afterwards, while no significant decrease
is seen above 30 keV. The QPO centroid frequencies are also related to photon energy.
Unlike the rms, it does not have a unified trend. In the four panels of Fig.\ref{fig:f6} (the top two of each column),
with the increasing of photon energy, the frequencies first increase and then decrease after $\sim$10 keV.
In the bottom two panels of the first column, the frequency is almost constant and independent with photon energy.
However, it shows monotonically increasing trend with photon energy for the rest of the panels.

\subsection{Phase Lags} \label{subsec:lag}

\begin{figure*}
\plottwo{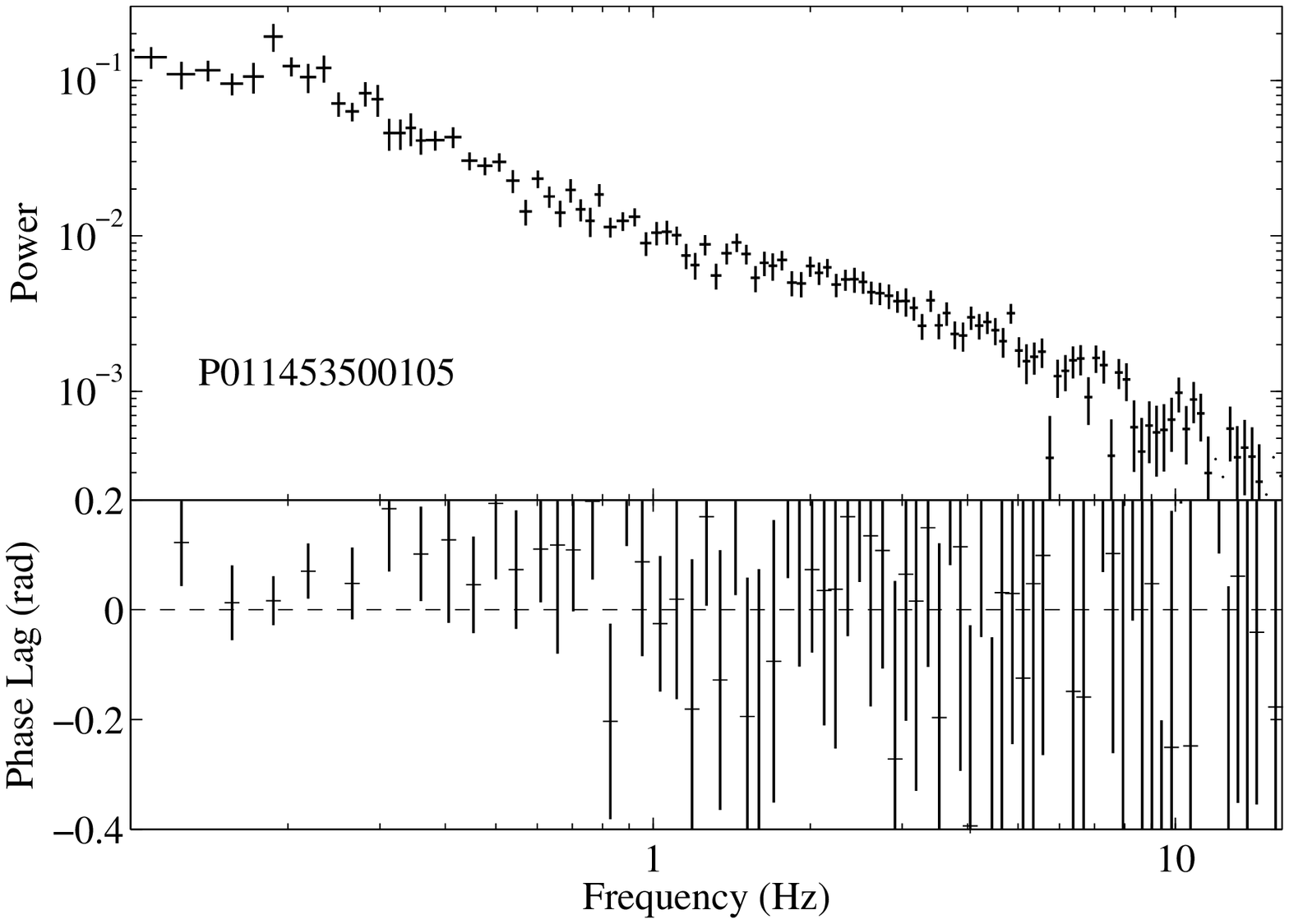}{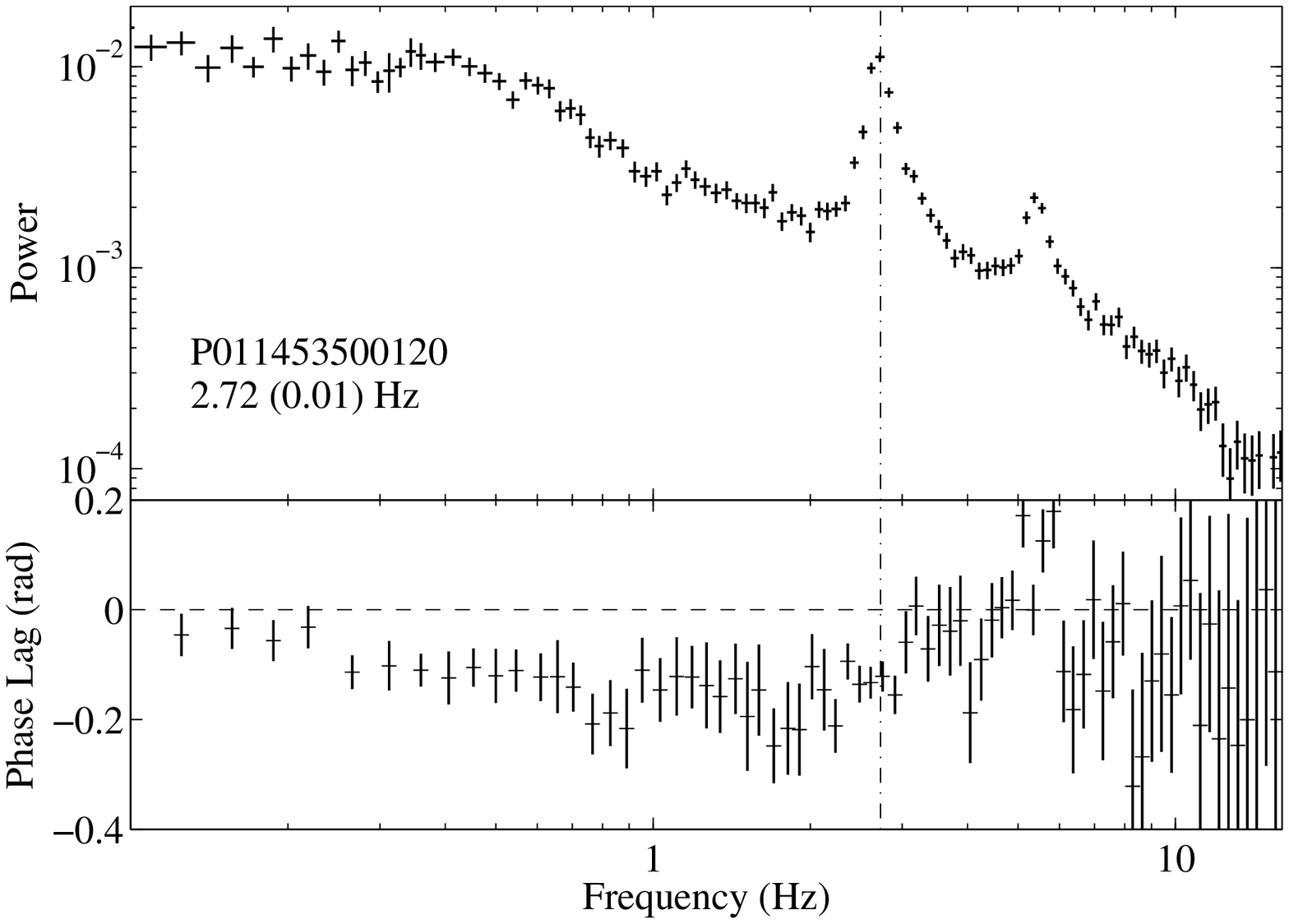}
\caption{Power density spectrum (upper) and phase lag spectrum (lower) of LE data sets
during the LHS (left) and the HIMS (right). The phase lags are calculated between the
light curves corresponding to 1-3 keV and 3-7 keV energy ranges.
The dashed vertical line marks the frequency of the QPO. \label{fig:f9}}
\end{figure*}

\begin{figure}
\plotone{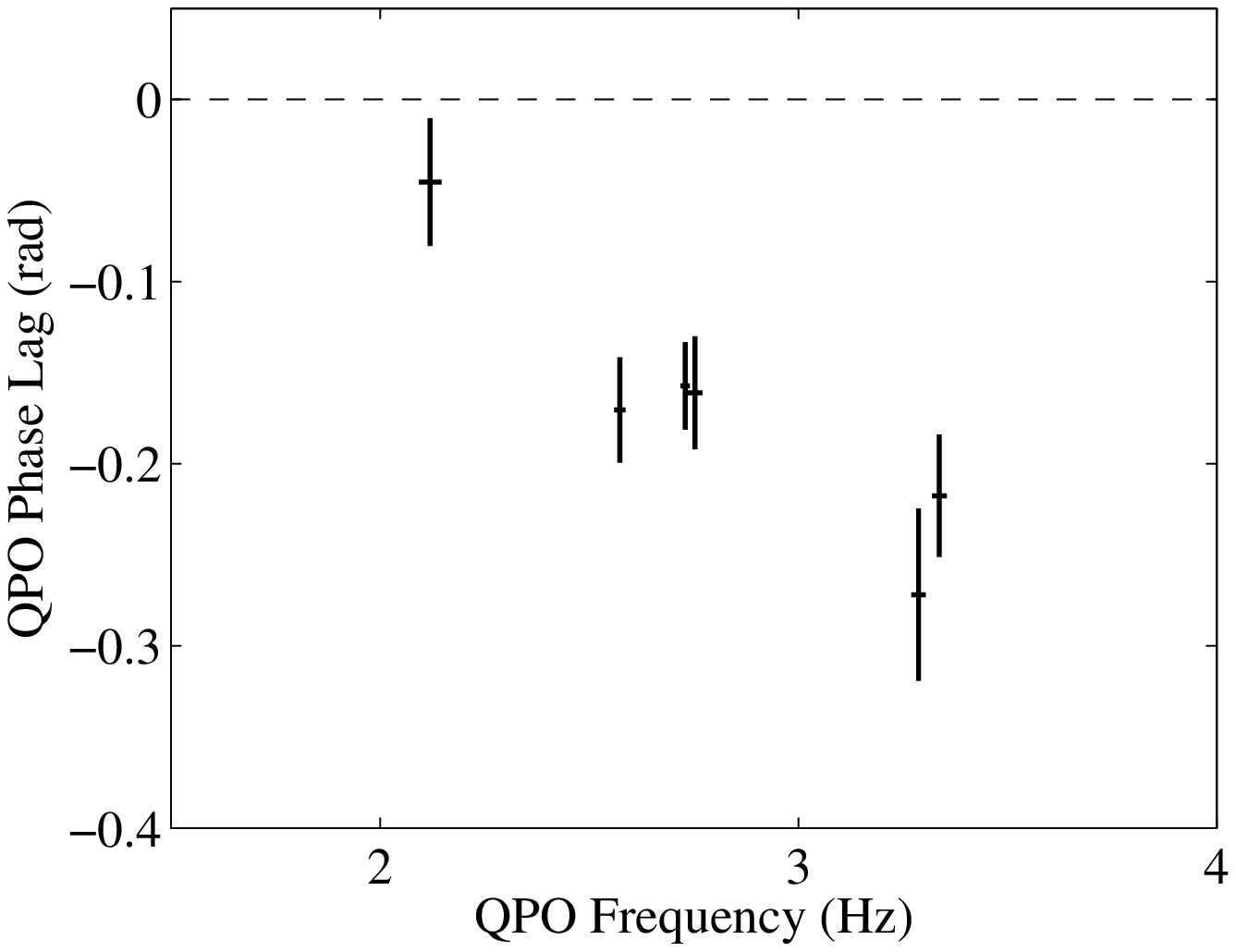}
\caption{Typc-C fundamental QPO phase lags as a function of the QPO frequency. \label{fig:f10}}
\end{figure}

Phase lags between soft and hard variabilities are computed from the LE data.
Due to the statistics limit, only the first period is selected for further study.
Fig.\ref{fig:f9} shows the phase-lags of two observations as a function of frequency
representing the LHS and the HIMS. Due to poor statistics, lags became hard to measure
at high frequencies; thus we plot them only below 16 Hz. In the LHS, the broad band noise
component shows a positive phase lag. During the HIMS, the lags of fundamental QPO are
negative, while the second harmonic shows positive phase lags. We also derive phase
lags at the QPO centroid frequency shown in Fig.\ref{fig:f10}.
The lag is strongly correlated with centroid frequency, with a trend towards zero lags while QPO frequency decreases.

\section{DISCUSSIONS} \label{sec:discussion}

\subsection{The outburst and source states} \label{subsec:d1}

In this work, we have presented the timing results of a new BHC MAXI J1535-571 during its outburst
in 2017 using \emph{Insight}-HXMT data. The outburst evolution is consistent with the scenario
typically observed in BHCs \citep{2005A&A...440..207B, 2011MNRAS.415..292M, Belloni2010}.
Based on the combined timing and color properties, we have identified three main states according
to the classification criteria given by Belloni \citeyear{Belloni2010}. The source experienced
a state transition from the LHS to HIMS in the early phase and then to SIMS.

Fig.\ref{fig:f2} shows typical hardness and timing properties of the canonical LHS, although
the vertical branch of the HID is completely observed. The PDS is dominated by a strong band-limited
noise (see panel (a) in Fig.\ref{fig:f4}), with typical rms values of $\sim$26\%. From observation P011453500119 (MJD 58008), the hardness
ratio shows a significant decline from $\sim$2.1 to $\sim$1.5 before the source enters the top left
in the HID (see Fig.\ref{fig:f2}). During this period, the PDS (Fig.\ref{fig:f4} panel (b) and (c)) \red{showed} a
band-limited noise and a strong typc-C QPO with a comparable lower total rms than in the LHS.
The results indicate the source enters the HIMS. Around MJD 58015 the timing variabilities show a
clear difference from those in the HIMS, while a type-B QPO appears in the PDS
(Fig.\ref{fig:f4} panel (d) and Fig.\ref{fig:f12}), indicating a transition to SIMS.

The above state transition are also consistent with the spectral fit of the \emph{Swift}
observations given by \citet{2017submit}. As shown in their paper,
the power law photon index $\Gamma$ stayed around $\sim$1.5 until MJD 58007, and increased suddenly
to $\sim$2.0 until MJD 58014. From MJD 58015, $\Gamma$ increased from $\sim$2.0 to $\sim$2.5.
The inner disk temperature and the disk flux ratio stabilized at a low value before MJD 58015,
and jumped to a high value afterward.

\subsection{Quasi-periodic oscillations} \label{subsec:d4}

The LFQPOs, consisting of three types (type-A -B and -C), are observed in the range 1.78 to 13.88 Hz (see Table \ref{tab:table2}).

Type-C QPOs are observed in HIMS, similar to XTE J1859+226 \citep{2004A&A...426..587C}, and in the early stages of SIMS,
with a higher centroid frequency, by all the three detectors. When type-C QPOs are observed in SIMS, a hard flaring
happened, suggesting an association to the hard component. Their frequencies are correlated with count rates and
hardness, similar to what have been observed in other BHT \citep{2001ApJ...548..401T,2005A&A...440..207B}.
The QPO frequencies observed by \emph{Insight}-HXMT are consistent with the \emph{NICER} results which showed QPO
frequency between 1.9 and 2.8 Hz during September 12, 10:53:39 and September 13, 22:40:40 \citep{2017ATel10768....1M}.
In our case, the QPO frequency is between 1.78 and 2.74 Hz during September 12, 10:38:59 and September 14, 08:06:59.

\begin{figure}
\plotone{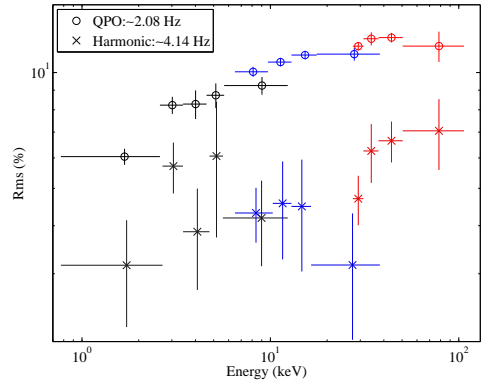}
\caption{Fractional rms spectra at the fundamental QPO frequency and its harmonic. \label{fig:f13}}
\end{figure}

The second harmonics of type-C QPOs are constantly detected in LE and HE observations, but only in some of the ME observations, which might
due be to the low signal-to-noise ratio (see Fig.\ref{fig:f3} and Fig.\ref{fig:f5}). For observation P011453500301,
the second harmonic is clearly detected in ME energy band, thus we can measure the fractional rms as a function of
photon energy for both QPO and its second harmonic (see Fig.\ref{fig:f13}). The relation of the rms of
the second harmonic QPOs with photon energy has been observed in XTE J1550-564 \citep{2013MNRAS.428.1704L} and GRS 1915+105 \citep{2016ApJ...833...27Y}.
However, while it displays an arch-like relation and the maximum amplitude of the arch relation appears at $\sim$7 keV
in XTE J1550-564, the rms of the harmonic QPO increases till $\sim$10 keV and then seems to decline until $\sim$30 keV
with large uncertainty in GRS 1915+105. Using frequency-resolved spectroscopy, \citet{2016MNRAS.458.1778A} found that
the second harmonic spectrum is dramatically softer than the QPO spectrum and the time-averaged spectrum, and can be
described by an additional soft Comptonization component. The lack of the second harmonic in ME observations may be due to a
physical reason. However, beyond $\sim$30 keV the fractional rms of the second harmonic increases with photon energy,
suggesting that the second harmonic may be also related to an additional component (i.e.,the reflection component).

Type-B QPOs are usually detected when a source experienced a rapid transition to the SIMS.
Fast transitions have been observed in GS 1124-68 \citep{1997ApJ...489..272T},
XTE J1859+226 \citep{2004A&A...426..587C} and GX 339-4 \citep{2005A&A...440..207B}.
A very sharp threshold in count rate was observed, suggesting a transition.
However, for MAXI J1535-571 the QPO has a frequency around $\sim$10 Hz, which is different from
the typical frequency of $\sim$6 Hz. The correlation of type-B QPO frequencies with the power-law flux
has been reported by \citet{2011MNRAS.418.2292M} and \citet{2017MNRAS.466..564G}.
The higher frequency of the type-B QPO could indicate
that MAXI J1535-571 has a higher hard luminosity compared to other systems.
Jet ejections are thought to be associated with Type-B QPOs
and the X-ray flux peak \citep{2009MNRAS.396.1370F}. In MAXI J1535-571, type-B QPO is found
in correspondence with the count rate peak (see Fig.\ref{fig:f1}). Future multi-wavelength observation are
needed to verify the existence of relativistic jet emission during the X-ray flux peak.

Type-A QPOs are observed in SIMS, with a clear QPO peak at around 10 Hz present only in ME and HE observations.
A similar behavior has been reported in GX 339-4 \citep{2005A&A...440..207B}.

\subsection{Energy dependence of QPO parameters } \label{subsec:d2}

For the first time we studied the fractional rms and the centroid
frequency of the QPO as a function of photon energy up to 100 keV(see Fig.\ref{fig:f8} and Fig.\ref{fig:f6}).

The QPO rms amplitude increases with photon energy till $\sim$20 keV and keeps more or less as a constant
in all the observations. The background estimation we applied is based on the blind FoV detectors.
The background consists of Cosmic X-ray and Particle Background including cosmic rays,
albedo radiation and SAA-induced background for a Low-Earth Orbit satellite \citep{2015Ap&SS.360...47X}.
Since there is no sign of other bright sources in the \emph{MAXI} images\footnote{\url{http://maxi.riken.jp/star\_data/J1535-572/J1535-572.html}},
most of the LE detector background comes from cosmic x-rays background (dozens counts s$^{-1}$),
which can be neglected compared to the high count rate in MAXI J1535-571. However, for ME and HE,
the background is dominated by Particle Background, which is related to the position and attitude of the satellite.
The HE and ME background typically accounts for $\sim$10\% to $\sim$20\% for sub-energy bands, except for the highest
sub-energy band of HE detector which can be around 50$\%$. In order to investigate the accuracy of our background estimation,
we applied several blank sky observations, and found the count rate ratio between the small FoV and the blind FoV
detectors is independent of time. Our background estimation method is thus reasonable for rms calculation.

In addition to MAXI J1535-571, similar energy dependence relations for the type-C QPO were found
in GRS 1915+015 \citep{2004ApJ...615..416R,2012Ap&SS.337..137Y,2013MNRAS.434...59Y,2016ApJ...833...27Y},
H1743-322 \citep{2013MNRAS.433..412L}, XTE J1859+226 \citep{2004A&A...426..587C} and XTE J1550-564 \citep{2013MNRAS.428.1704L},
in which a corona origin of type-C QPOs is considered. For GRS 1915+105,
HEXTE results showed that the QPO rms decreases above 20 keV \citep{2001ApJ...548..401T}.
However, \citet{2004ApJ...615..416R} found that this cut-off was
not always present, but rather related to the compact jets which contributes to the hard X-ray
component mostly through synchrotron emission.
\citet{2018ApJ...858...82Y} computed the fractional rms spectrum of the QPO
in the context of the Lense$-$Thirring precession model \citep{2009MNRAS.397L.101I}. They found that the rms at higher energy
$E > 10$ keV becomes flat when the system being viewed with large inclination angle.
Our result is consistent with the simulation.

The correlation between the centroid frequency of QPOs and the photon energy shows three different shapes:
flat, positive and `arch' like. For energies $<20$ keV, this relation in GRS 1915+105 \citep{2010ApJ...710..836Q,2012Ap&SS.337..137Y,2018MNRAS.474.1214Y}
and XTE J1550-564 \citep{2013MNRAS.428.1704L} evolves from a negative correlation to a positive one when the QPO
frequency increases, but with a different turn-over QPO frequency. The pattern in H1743-322 shows no apparent
turn-over frequency, which might be due to the lack of observational data for the hard state \citep{2013MNRAS.433..412L}.
The energy dependence of the QPO frequency could be caused by differential precession of the inner accretion flow \citep{2016MNRAS.458.3655V}.
The inner$-$part flow causes a higher QPO frequency than the outer$-$part flow, and the evolution of the spectral properties of the
inner and outer part can causes the frequency$-$energy relation change from negative to positive.
When the inner$-$part flow has a harder spectrum than the outer$-$part flow, this causes a positive correlation.
In MAXI J1535-571, the turn over of the relation at high energy $E > 10$ keV would suggest that
it is due to the reflection bump being prominent at those energies. The reflected spectrum is expected to
be dominated by photons emitted by the outer$-$part flow, thus the reflected spectrum will show a
relatively low precession frequency.

\subsection{Phase lag and Inclination estimates} \label{subsec:d3}

We have calculated the phase lag between the 1-3 keV and 3-7 keV energy bands.
We have found that the phase lags of the fundamental and the harmonic of type-C QPOs keeps opposite.
The lags of the fundamental peak are soft, while the harmonic show hard lags.
Similar to that found in GRS 1915+105 \citep{2000ApJ...543L.141L, 2000ApJ...541..883R, 2010ApJ...710..836Q},
and XTE J1859+226 \citep{2004A&A...426..587C}, the lag is strongly correlated with the centroid frequency of the QPO,
and decreases with an increasing frequency.

Recently, from the inclination dependence of phase lags in a sample of 15 black hole binaries,
\citet{2017MNRAS.464.2643V} found that the phase lag of the type-C QPOs
strongly depends on the inclination, both in evolution with the QPO frequency and sign. All samples possess
a slightly hard lag at low QPO frequencies. At high frequencies high-inclination sources turn to soft lags
while lags in low-inclination sources become harder. These results support the geometrical origin of type-C QPOs.

MAXI J1535-571 clearly follows the trend of high-inclination sources presented in
\citet{2017MNRAS.464.2643V}. \citet{2018ApJ...852L..34X}
performed a spectral analysis of the \emph{NuSTAR} observation in the hard state,
and found that the energy spectra can be well fitted by two different models which both consist of a multi-temperature thermal component,
but with different reflection models (one for {\tt\string relxilllpCp+xillverCp}, the other for {\tt\string relxillCp+xillverCp}).
They found that the inclination angle is $57_{-2}^{+1}$$^{\circ}$ or $75_{-4}^{+2}$$^{\circ}$, respectively.
\red{And, the spectral fitting result from \emph{NICER} suggested a similar inclination of $67.4(8)$$^{\circ}$ \citep{2018ApJ...860L..28M}.
Both are consistent with our phase lags result}.

\section{CONCLUSION} \label{sec:conclusion}

We have presented timing analysis of the new BHC MAXI J1535-571 using \emph{Insight}-HXMT observations.
The main results of the study are:

1) The source exhibits state transitions from LHS to HIMS, and then SIMS.

2) For the first time an energy dependence of the QPO fractional rms and frequency is observed up to 100 keV.
While the energy dependence rms is consistent with other black hole binaries observed by RXTE,
\emph{Insight}-HXMT reveals that the frequency-energy relation changes dramatically.

3) By assuming a geometric origin of type-C QPO, MAXI J1535-571 is consistent with being a high inclination source.

\acknowledgements {

This work made use of the data from the Insight-HXMT mission, a project funded by China National Space Administration (CNSA)
and the Chinese Academy of Sciences (CAS). The Insight-HXMT team gratefully acknowledges the support from
the National Program on Key Research and Development Project (Grant No. 2016YFA0400800).
J. L. Qu acknowledges the National Science Foundation of China 1173309 and 11673023.

}

\bibliography{biblio}

\end{document}